\documentclass[12pt,a4paper]{revtex4}
\usepackage{graphicx}
\usepackage{xcolor}
\usepackage{tabularx}
\usepackage{hyperref}
\usepackage{fancybox}
\usepackage{slashed}
\usepackage{amsfonts}
\usepackage{graphicx} 
\usepackage{pdfpages}
\usepackage{amsmath} 
\usepackage{empheq}
\usepackage{bbold}
\usepackage{amssymb} 
\usepackage{epstopdf} 
\usepackage[utf8]{inputenc}       
\usepackage[T1]{fontenc} 
\usepackage{feynmf}
\usepackage{mathtools}
\usepackage{upgreek}
\usepackage{tabularx} 
\usepackage{subfig}
\usepackage{psfrag} 
\usepackage{sistyle} 
\usepackage{ytableau}
\usepackage{youngtab}
\usepackage{color} 
\begin{document}
\title{Large $N_f$ for multiple representations}

\author{Giacomo Cacciapaglia}
\email{g.cacciapaglia@ipnl.in2p3.fr}
\affiliation{Univ. Lyon, Universit{\'e} Claude Bernard Lyon 1, CNRS/IN2P3, UMR5822 IP2I, F-69622, Villeurbanne, France} 
\author{Shahram Vatani}
\email{vatani@ipnl.in2p3.fr}
\affiliation{Univ. Lyon, Universit{\'e} Claude Bernard Lyon 1, CNRS/IN2P3, UMR5822 IP2I, F-69622, Villeurbanne, France} 

\begin{abstract}
We present an extension of the large $N_f$ formalism that allows to study cases with multiple fermion representations. The pole structure in the beta function is traced back to the intrinsic non-abelian nature of the gauge group, independently on the fermion representation. This result validates the conjectured existence of an interactive UV fixed point for non-abelian gauge theories with large fermion multiplicity. Finally, we apply our results to chiral gauge theories and to extended Grand Unified Theories.
 \\[.1cm]
 \end{abstract}

\maketitle

\section{Introduction}

The expansion in large number of fermionic matter fields, $N_f$, has been used to study the dynamical properties of gauge theories. In particular, the gauge beta function, relevant for the renormalisation group equation (RGE) of the gauge coupling and the mass anomalous dimension have been computed for both abelian~\cite{Espriu:1982pb,PalanquesMestre:1983zy} and non abelian~\cite{Gracey:1996he} theories. Some information about higher-order terms is also available~\cite{Ciuchini:1999cv,Ciuchini:1999wy,Dondi:2020qfj}. More recently, this technique has been reused to show that gauge theories in the large-$N_f$ limit may feature a non-trivial, interacting Ultra-Violet (UV) fixed point~\cite{Antipin:2017ebo}. This observation is very important in understanding the dynamics of gauge theories, as the presence of an UV fixed point would allow to understand large-$N_f$ gauge theories as fundamental, in the Wilsonian sense~\cite{Wilson:1971bg,Wilson:1971dh}.  This effort falls in the larger quest for asymptotic freedom, first identified by S.Weinberg for quantum gravity~\cite{Weinberg:1980gg} and later discovered by F.Sannino and D.Litim for perturbative gauge-Yukawa theories~\cite{Litim:2014uca}.

The presence of a fixed point is linked to the fact that the first order in the large-$N_f$ expansion has a negative pole at a given value of the gauge coupling, thus cancelling the positive leading term when the gauge coupling grows near the singular value.  This conclusion has been recently challenged in Ref.~\cite{Alanne:2019vuk}, where the resummation was re-organised thanks to non-perturbative information obtained from critical exponents. The result seems to imply that a singularity in the critical exponent does not necessarily imply a singularity in the beta function, however the analysis is based on assumptions that do not allow to draw reliable conclusions~\cite{Sannino:2019vuf}. Preliminary results from large-$N_f$ lattice studies also remain inconclusive~\cite{Leino:2019qwk}. Thus, at the moment the presence of a physical UV fixed point cannot be excluded. This phenomenon has been applied in various contexts, from attempts to define a safe Standard Model~\cite{Abel:2017rwl,Mann:2017wzh,Pelaggi:2017abg,Abel:2018fls}, to Grand Unification~\cite{Molinaro:2018kjz,Wang:2018yer}, Dark Matter~\cite{Sannino:2014lxa,Cai:2019mtu} and a variety of New Physics scenarios~\cite{Cacciapaglia:2018avr,Sannino:2019sch,Ryttov:2019aux}. In composite Higgs models with fermion partial compositeness~\cite{Cacciapaglia:2018avr}, two \emph{irreps} are needed to form spin-1/2 bound states that couple to the elementary quark and lepton fields.

The basic formulae for large-$N_f$ resummation are known~\cite{Holdom:2010qs} for a simple gauge group $\mathcal{G}$, while an extension to semi-simple groups $\mathcal{G}=\times_\alpha \mathcal{G}_\alpha$ can be found in Ref.~\cite{Antipin:2018zdg}. In all cases, the large multiplicity fermions belong to a single irreducible representation (\emph{irrep}) $R_f$ of the gauge groups. 
In this note, we generalise the resummation to cases with a large multiplicity of fermions in multiple \emph{irreps} of the gauge groups. We find that the pole structure is preserved, revealing that its presence is intrinsically linked to the non-abelian structure of the gauge group. This result is in agreement with the presence of the UV safety for non-abelian gauge groups and not for abelian ones. The latter was already in question due to the fact that the mass anomalous dimension diverges near the pole~\cite{Antipin:2017ebo}. Our results are also relevant to understand the dynamical properties of some class of models, like gauge theories with a large number of chiral families, and composite Higgs models with top partial compositeness~\cite{Cacciapaglia:2018avr}.

The paper is organised as follows: in Sec.~\ref{sec2} we review the main results useful for large-$N_f$ resummation, while presenting the results in a different form that can be applied to the case of multiple \emph{irreps}. In Sec.~\ref{sec3} we present general formulae for the new case, before applying the results to physically interesting theories in Sec.~\ref{sec4}. We offer our conclusion in Sec.~\ref{sec5}.

\section{Basic resummation results} \label{sec2}

In this section we will give a pedagogical introduction to the basic results for the large-$N_f$ beta-function calculation, following Ref.~\cite{Antipin:2018zdg}. We will however change some definitions, which will be useful to better understand the origin of the singularity and to extend the calculation to multiple \emph{irrep} cases in the next section.

\begin{figure}[h]
    \centering
\begin{fmfgraph}(200,50)
    \fmfleft{i}
    \fmfright{o}
    \fmfpolyn{fermion,smooth,tension=0.58}{v}{4}
     \fmfpolyn{fermion,smooth,tension=0.58}{u}{4}
     \fmfpolyn{fermion,smooth,tension=0.58}{w}{4}
     \fmfpolyn{fermion,smooth,tension=0.58}{x}{4}
     \fmfpolyn{fermion,smooth,tension=0.58}{y}{4}
    \fmf{curly}{i,v1}
    \fmf{curly}{v3,u1}
    \fmf{curly}{u3,w1}
     \fmf{curly}{w3,x1}
      \fmf{curly}{x3,y1}
       \fmf{curly}{y3,o}
\end{fmfgraph}
   \caption{ Example of \textit{bubble-chain} of length $5$.}
    \label{fig:bubblechain}
\end{figure}
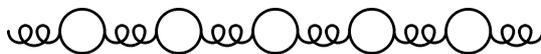

Let us consider a gauge-fermion theory with one species of Dirac fermions, $\Psi$, in the \emph{irrep} $R$ of a simple gauge group $\mathcal{G}$ (with gauge coupling constant $g$). To compute the $\beta$-function we need to calculate the radiative corrections to the 2-point function of the gauge boson propagator. For large number of fermions, the leading contribution to the beta function comes from the one loop fermion contribution:
\begin{fmffile}{diagram}
\begin{equation}
    \begin{gathered}
     \begin{fmfgraph*}(100,50)
     \fmfleft{i} 
     \fmfright{o} \fmfpolyn{fermion,smooth,tension=0.58}{v}{4} \fmf{curly}{i,v1} 
     \fmf{curly}{v3,o}
     \end{fmfgraph*} 
    \end{gathered}\;\;\;=\;\; K \; \times  \int \frac{\mathrm{d}^4 k}{4\pi^2} \frac{\mathrm{Tr}\left[ \gamma^\mu \left( \slashed{k} - \slashed{p}\right) \gamma^\nu \slashed{k} \right]}{\left(p-k\right)^2 k^2} \,,
\label{eq:1loop}
\end{equation}
where we have defined an effective gauge coupling $K$, which takes into account the large fermion multiplicity as
\begin{equation}
K = \frac{g^2}{4 \pi^2} N_f\ T(R)\,,
\end{equation}
where $T(R)$ is the index of the fermion \emph{irrep} $R$. $K$ can be considered as the effective coupling controlling the perturbative expansion, thus the one-loop contribution can be counted at $\mathcal{O} (N_f^0)$. This also allows to define a chain of fermion bubbles, which are all contributing at $\mathcal{O} (N_f^0)$, as shown in Fig.~\ref{fig:bubblechain}.

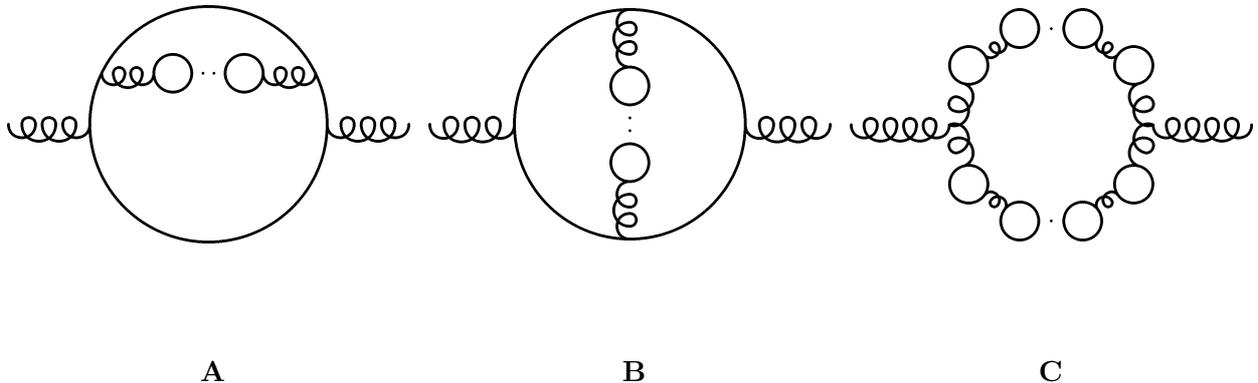
\begin{figure}[tb]
\begin{tabular}{ccc}
\subfloat{\begin{fmfgraph*}(150,150)
    \fmfleft{i}
    \fmfright{o}
    \fmfrpolyn{fermion,smooth,tension=0.6}{v}{14}
    \fmfv{decor.shape=circle,decor.filled=empty}{u1,u2}
    \fmf{gluon}{v2,u1}
    \fmf{dots}{u1,u2}
    \fmf{gluon}{u2,v7}
    \fmf{curly}{i,v1}
    \fmf{curly}{v8,o}
    \fmf{phantom,label=, label.side=right}{i,v1}
    \fmf{phantom,label=, label.side=left}{v8,o}
    \fmfiv{l=\LARGE\textbf{A},l.a=-155,l.d=0.4w}{c}
    \end{fmfgraph*}
    }
&
\subfloat{\begin{fmfgraph*}(150,150)
    \fmfleft{i}
    \fmfright{o}
    \fmfrpolyn{fermion,smooth,tension=0.2}{v}{4}
     \fmfv{decor.shape=circle,decor.filled=empty}{u1,u2}
     \fmf{gluon}{v2,u1}
     \fmf{dots}{u1,u2}
     \fmf{gluon}{u2,v4}
    \fmf{curly}{i,v1}
    \fmf{curly}{v3,o}
    \fmf{phantom,label=, label.side=right}{i,v1}
    \fmf{phantom,label=, label.side=right}{v3,o}
    \fmfiv{l=\LARGE\textbf{B},l.a=-155,l.d=0.4w}{c}
\end{fmfgraph*}
    }   
&
\subfloat{\begin{fmfgraph*}(150,150) 
    \fmfleft{i}
    \fmfright{o}
    \fmfrpolyn{phantom,smooth,tension=0.01}{v}{10}
    \fmfv{decor.shape=circle,decor.filled=empty}{v2,v3,v4,v5,v7,v8,v9,v10}
    \fmf{curly}{v1,v2}
    \fmf{curly}{v2,v3}
    \fmf{dots}{v3,v4}
    \fmf{curly}{v4,v5}
    \fmf{curly}{v5,v6}
    \fmf{curly}{v6,v7}
    \fmf{curly}{v7,v8}
    \fmf{dots}{v8,v9}
    \fmf{curly}{v9,v10}
    \fmf{curly}{v10,v1}
    \fmf{curly}{i,v1}
    \fmf{curly}{v6,o}
    \fmf{phantom,label=, label.side=left}{i,v1}
    \fmf{phantom,label=, label.side=left}{v6,o}
    \fmfiv{l=\LARGE\textbf{C},l.a=-155,l.d=0.4w}{c}
\end{fmfgraph*}} \\
{\bf A} & {\bf B} & {\bf C}
\end{tabular}
\caption{Next-to-leading order diagrams. {\bf C} is one representative of the diagrams containing gauge boson self-interactions.} \label{fig:NLO}
\end{figure}

Two-loop corrections to the diagram in Eq.~\eqref{eq:1loop} correspond to attaching a gauge propagator to the fermion loop: as no additional $N_f$ multiplicity is added, this diagram will effectively contribute to order $K^2/N_f$, thus providing next-to-leading order terms. Now, replacing the simple gauge propagator with a \textit{bubble-chain}, will not increase the $1/N_f$ order. In fact, the leading term is simply given by the resummation of the \textit{bubble-chain} in the gauge propagator, as shown in the first two diagrams in Fig.~\ref{fig:NLO}. For a non-abelian theory, there are also contributions coming from gauge boson self-interactions: in this case, even the one-loop result is suppressed by $1/N_f$ and should be considered at next-to-leading order. As before, the resummed results stem from dressing the gauge propagators with the \textit{bubble-chain}, as exemplified in the third diagram in Fig.~\ref{fig:NLO}.

For each type of diagram \textbf{X}, we can write the amplitude in the form $\delta^{ab} p^2 \Updelta_{\mu \nu} \left( p \right) \Uppi_{\textbf{X}}^{\left(n\right)}\left(p \right)$, where  $ \Updelta_{\mu \nu}(p)=\eta_{\mu \nu} - p_\mu p_\nu /p^2$ . Here $n$ corresponds to the length of the \textit{bubble-chain} in \textbf{A} and \textbf{B}, while for \textbf{C} it's the total length of the two \textit{bubble-chains}.
A simple calculation gives:
\begin{eqnarray}
\Uppi_{\textbf{A}}^{\left(n\right)}(p) \enskip  & = & N_f g^4 \; T(R)\ C(R)\; \frac{K^{n}}{(4 \pi^2)^2} A^{\left(n\right)}_\textbf{A}(p)\,,\\
\Uppi_{\textbf{B}}^{\left(n\right)}(p) \enskip  & = & N_f g^4 \; T(R)\ C(R)\ \left( 1 - \frac{1}{2}\frac{C(G)}{C(R)}\right)\; \frac{K^{n}}{(4\pi^2)^2} A^{\left(n\right)}_\textbf{B}(p)\,, \\
\Uppi_{\textbf{C}}^{\left(n\right)}(p) \enskip  & = & g^2 \; C(G)\; \frac{K^{n}}{(4 \pi^2)} A^{\left(n\right)}_\textbf{C}(p)\,,
\end{eqnarray}
where $C(R)$ the Casimir of the \emph{irrep} $R$ and $C(G)$ of the adjoint. 
The functions $A_{\textbf{X}}^{(n)}$ are integrals of the loop momentum and contain the information needed to compute the leading order, however we need to take into account all the $n$-long \textit{bubble-chains} to extract the contribution to the $\beta$-function. 
The beauty of the large-$N_f$ expansion stands in the fact that this resummation can be done, and the $\epsilon$-dependence of the loop in dimensional regularisation can be converted in a dependence on $K$ of the resummed result (see Ref.~\cite{PalanquesMestre:1983zy} and the appendix of Ref.~\cite{Antipin:2018zdg} for the proof).

To compute the evolution of the coupling we need to sum up all the diagrams: the total contribution thus reads
\begin{multline}
\Uppi=\sum_n 2\Uppi_{\textbf{A}}^{\left(n\right)} + \Uppi_{\textbf{B}}^{\left(n\right)} + \Uppi_{\textbf{C}}^{\left(n\right)}  =   \frac{K}{N_f T(R)} C( G)\ \sum_n K^n A^{(n)}_{\textbf{C}}(p)  + \\
 \enskip \frac{K^2}{N_f T(R)} C(R)\ \sum_n K^{n} \;\left[  2 \; A^{\left(n\right)}_\textbf{A}\left(p\right) + \left( 1 - \frac{1}{2}\frac{C \left(G \right)}{C\left(R \right)}\right)  A^{\left(n\right)}_\textbf{B}\left(p\right) \right] \,,
\end{multline}
where the factor of $2$ comes from the two possible insertions of the \textit{bubble-chain} in the diagram \textbf{A}. 
Upon closer inspection to the above formula, we can reorganise the sum as follows:
\begin{equation}
\Uppi= \enskip  \frac{K}{N_f T(R)}  \sum_{n} \left\{   C(R)\ K^{n}\underbrace{\left[ 2  A_\textbf{A}^{\left(n-1\right)} + A_\textbf{B}^{\left(n-1\right)} \right]}_{(*)}   +  C(G)\  K^{n-1} \underbrace{\left[  A_\textbf{C}^{\left(n-1\right)} - K A_\textbf{B}^{\left(n-1\right)}/2 \right]}_{(**)} \right\}\,.
\end{equation}
The combination $(*)$ encodes the contribution to the beta function for an abelian gauge group (for which $C(G)=0$ and $C(R) \to Q_f^2$) and was computed originally in Ref.~\cite{PalanquesMestre:1983zy}, while the combination $(**)$ encodes the effect of the non-abelian dynamics. After resummation, we define two functions corresponding to the two combinations as follows
\begin{equation}
    \beta\left( K \right)= \frac{2K^2}{3} \left[ 1 + \frac{C\left( G \right)}{N_f T\left( R \right)}\left\{ \frac{-11}{4} + H\left(K\right)\right\} + \frac{C\left(R \right)}{N_f T\left(R \right)} F(K) \right]\,,
\label{totaleq}
\end{equation}
where $F(K)$ stems from $(*)$ and $H(K)$ from $(**)$. Note that the $-11/4$ term isolates the 1-loop contribution of gauge couplings, which is of order $1/N_f$, wile the $1$ corresponds to the one-loop contribution of the fermions. Thus, in our definition, the functions $F(K)$ and $H(K)$ explicitly contain only the resummed higher-loop contribution.
They are defined as~\footnote{The functions $F$ and $G$ we define are related to the $F_1$ and $H_1$ functions of Ref.~\cite{Antipin:2018zdg} as
$$
F_1 = F\,, \qquad H_1 = F_1 + \frac{C(G)}{C(R)}  \left( -\frac{11}{4} + H \right)\,.
$$ }:
\begin{equation}
    F(K) = \frac{3}{4} \int_0^K \mathrm{d}x\; \Tilde{F}\left(0,\frac{2}{3}x\right)\,, \qquad
    H(K) =\frac{3}{4} \int_0^K \mathrm{d}x\; \Tilde{F}\left(0,\frac{2}{3}x\right)\ \Tilde{G}\left(0,\frac{1}{3}x\right)\,;
    \label{eq:FH}
\end{equation}
with
\begin{eqnarray}
     \Tilde{F}\left(0,\epsilon\right) &=& \frac{\left( 1-\epsilon\right)\left( 1-\frac{\epsilon}{3}\right)\left( 1+\frac{\epsilon}{2}\right)\Gamma \left( 4-\epsilon\right)}{2\Gamma^2\left( 2-\frac{\epsilon}{2}\right)\Gamma\left( 3-\frac{\epsilon}{2}\right)\Gamma\left( 1+\frac{\epsilon}{2}\right)} \,, \\
     \Tilde{G} \left(0,\epsilon \right) &=& \frac{20-43 \epsilon+32 \epsilon^2-14 \epsilon^3+4 \epsilon^4}{4\left(2 \epsilon-1\right)\left(2 \epsilon-3\right)\left(1-\epsilon^2\right)}\,.
\end{eqnarray}

The function $F(K)$ has a singularity at $K^\ast = 15/2$: this specifically arises from the singularity in the factor $\Gamma \left( 4-\epsilon \right)$ in the loop integral $\Tilde{F} (0,\epsilon)$. This term is relevant for abelian gauge groups. Instead, $H(K)$ has a singularity at $K^\ast = 3$, which stems from the $(1-\epsilon^2)$ factor in the denominator of $\Tilde{G} (0,\epsilon)$. Thus, the presence of a pole at $K = 3$ for non-abelian gauge theories is to be traced back to the non-abelian nature of the gauge bosons. It has been observed in Ref.~\cite{Antipin:2018zdg} that the mass anomalous dimension is finite in $K^\ast = 3$, while it diverges at $K^\ast = 15/2$, thus supporting the presence of an UV fixed point for the non-abelian gauge only. Furthermore, preliminary results for the $1/N_f^2$ contribution to the abelian $\beta$-function show that a discontinuity in $K^\ast = 3$ emerges. Both observations seem to support the idea that the non-abelian fixed point may be physical, while the abelian one is more arguable. In the remainder of this work, we will therefore focus on non-abelian gauge symmetries.

\begin{figure}[tb]
\begin{tabular}{ccc}
\subfloat{\begin{fmfgraph*}(150,150)
    \fmfleft{i}
    \fmfright{o}
    \fmfrpolyn{fermion,smooth,tension=0.6}{v}{14}
    \fmfv{decor.shape=circle,decor.filled=empty}{u1,u2}
    \fmf{photon}{v2,u1}
    \fmf{dots}{u1,u2}
    \fmf{photon}{u2,v7}
    \fmf{curly}{i,v1}
    \fmf{curly}{v8,o}
    \fmf{phantom,label=, label.side=right}{i,v1}
    \fmf{phantom,label=, label.side=left}{v8,o}
    \fmfiv{l=\LARGE\textbf{A},l.a=-155,l.d=0.4w}{c}
    \end{fmfgraph*}
    }
&
\subfloat{\begin{fmfgraph*}(150,150)
    \fmfleft{i}
    \fmfright{o}
    \fmfrpolyn{fermion,smooth,tension=0.2}{v}{4}
     \fmfv{decor.shape=circle,decor.filled=empty}{u1,u2}
     \fmf{photon}{v2,u1}
     \fmf{dots}{u1,u2}
     \fmf{photon}{u2,v4}
    \fmf{curly}{i,v1}
    \fmf{curly}{v3,o}
    \fmf{phantom,label=, label.side=right}{i,v1}
    \fmf{phantom,label=, label.side=right}{v3,o}
    \fmfiv{l=\LARGE\textbf{B},l.a=-155,l.d=0.4w}{c}
\end{fmfgraph*}
    }   
 \\
{\bf A'} & {\bf B'} &
\end{tabular}
\caption{Next-to-leading order diagrams for the mixed contributions, where the gauge bosons in the \textit{bubble-chain} and on the external legs belong to different gauge groups.} \label{fig:NLOmix}
\end{figure}
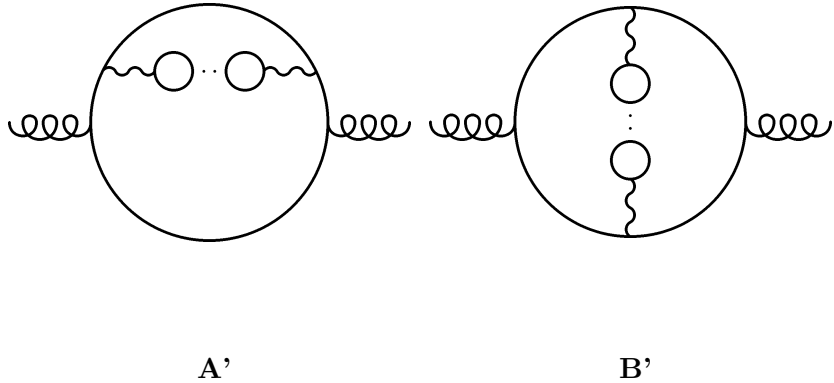

The resummation has been extended to semi-simple gauge groups in Ref.~\cite{Antipin:2018zdg}, and we will review the main results here.
We consider a gauge group $\mathcal{G} = \times_\alpha\mathcal{G}_\alpha$, with gauge couplings $g_\alpha$, and $n_f$ fermions in the \emph{irrep} $R_f = \times_\alpha R_\alpha$. Instead of defining a single effective $N_f$ as in Ref.~\cite{Antipin:2018zdg}, we find more convenient to define a fermion number $N_\alpha$ for each gauge group, as this will allow us to easily generalise the result to multiple \emph{irreps}. We will consider that each fermion number, defined as
\begin{equation}
N_\alpha = n_f \Pi_{\beta \neq \alpha} d(R_\beta)\,, 
\end{equation}
is of the same order for all gauge groups, i.e. $N_\alpha = \mathcal{O} (N_f)$. Similarly, we define effective gauge couplings as follows:
\begin{equation}
K_\alpha = \frac{g_\alpha^2}{4 \pi^2} N_\alpha\,.
\end{equation}
The only new ingredient in the case of semi-simple gauge groups is that the gauge couplings contribute to each other's $\beta$-function. As gauge bosons of different groups do not interact with each other, the leading order in $N_f$ stems from diagrams of type {\bf A} and {\bf B}, where the gauge boson in the \textit{bubble-chain} is different from the external ones, as shown in Fig.~\ref{fig:NLOmix}.
The amplitudes can be written as:
\begin{equation}
\Pi_{\textbf{A'/B'}}^{(n)} (p) = g_\alpha^2 T(R_\alpha)\, \sum_{\beta \neq \alpha} g_\beta^2 C(R_\beta) d(R_\beta) \Pi_{\gamma \neq \alpha, \beta} d(R_\gamma)\ n_f\; \frac{K_\beta^n}{(4 \pi^2)^2} A^{(n)}_{\textbf{A/B}} (p)\,,
\end{equation}
where the integral functions are the same as before. Using $d(R_\beta) \Pi_{\gamma \neq \alpha, \beta} d(R_\gamma)\ n_f = N_\beta$, the total contribution can be written as:
\begin{equation}
\Pi = \frac{K_\alpha}{N_\alpha} \sum_{\beta \neq \alpha} \frac{C(R_\beta)}{T(R_\beta)}\ K^{n+1}_\beta \left( 2 A_{\textbf{A}}^{(n)} (p) +A_{\textbf{B}}^{(n)} (p) \right) \,,
\end{equation}
where we recognise the function also appearing in the abelian case.

Finally, the $\beta$-function for $K_\alpha$ can be written as
\begin{equation}
\beta (K_\alpha) = \frac{2 K_\alpha^2}{3} \left[ 1 + \frac{C(G_\alpha)}{N_\alpha T(R_\alpha)} \left\{ -\frac{11}{4} + H(K_\alpha) \right\} + \sum_\beta \frac{C(R_\beta)}{N_\alpha T(R_\beta)} F(K_\beta) \right]\,.
\end{equation}
If we only consider non-abelian gauge groups, the second term, which includes the mixed contributions, will always remain finite and small as $K_\beta < 3$. Thus, the presence of an UV fixed point only comes from the first term, i.e. from the non-abelian nature of each gauge factor in the semi-simple group.

In order to offer a direct comparison with the results in the next section, we can also redefine the fermion multiplicities by absorbing the index of the \emph{irrep}:
\begin{equation}
\Tilde{N}_\alpha = N_\alpha T(R_\alpha)\,, \quad K_\alpha = \frac{g_\alpha^2}{4 \pi^2} \Tilde{N}_\alpha\,,
\end{equation}
with the $\beta$-function given by
\begin{equation}
\beta (K_\alpha) = \frac{2 K_\alpha^2}{3} \left[ 1 + \frac{C(G_\alpha)}{\Tilde{N}_\alpha} \left\{ -\frac{11}{4} + H(K_\alpha) \right\} + \frac{1}{\Tilde{N}_\alpha} \sum_\beta \frac{C(R_\beta) T(R_\alpha)}{T(R_\beta)} F(K_\beta) \right]\,.
\label{eq:beta_semisimple}
\end{equation}
In the above form, the $\beta$-function can be easily extended to cases with a large multiplicity of multiple \emph{irreps}, as we will show in the coming section.

\section{Extension to multiple \emph{irreps}} \label{sec3}

We are now ready to extend the resummation formulae to cases with large numbers of multiple \emph{irreps}. We will first start with a simple gauge group, and then generalise to semi-simple groups.

\subsection{Simple Gauge Group}

We start with the case of a simple gauge group $\mathcal{G}$ with $n_i$ fermions $\Psi_i$ in different \emph{irreps} $R_i$. 
In such a case, the leading order contribution is given by the one-loop diagram below, where we sum over all the fermion species:
\begin{equation}
    \begin{gathered}
     \begin{fmfgraph*}(100,50)
     \fmfleft{i} 
     \fmfright{o} \fmfpolyn{fermion,smooth,tension=0.58}{v}{4} \fmf{curly}{i,v1} 
     \fmf{curly}{v3,o}
     \end{fmfgraph*} 
    \end{gathered}\;\;\;=\;\; \frac{g^2}{4\pi^2}\sum_i T\left( R_i \right) n_i \; \times \int \frac{\mathrm{d}^4 k}{4\pi^2}\; \frac{\mathrm{Tr}\left[ \gamma^\mu \left( \slashed{k} - \slashed{p}\right) \gamma^\nu \slashed{k} \right]}{\left(p-k\right)^2 k^2}\,.
\end{equation}
Following Eq.~\eqref{eq:1loop}, we can define an effective gauge coupling as follows:
\begin{equation}
K=\frac{g^2}{4\pi^2} N\,, \qquad N = \sum_i n_i T(R_i)\,,
\label{eq:Ndef}
\end{equation}
where it is evident why we absorbed the index of the \emph{irrep} in the definition of the fermion multiplicity. 
Interestingly, a large $N$ can now be due also by the contribution of \emph{irreps} with large index.

The next-to-leading order in $1/N$ is given by the same diagrams \textbf{A}, \textbf{B} and \textbf{C} in Fig.~\ref{fig:NLO}, yielding:
\begin{eqnarray}
\Uppi_{\textbf{A}}^{(n)}(p) \enskip  & = & g^4 \; \sum_i T(R_i)\ C(R_i) n_i\; \frac{K^{n}}{(4 \pi^2)^2} A^{(n)}_\textbf{A}(p)\,, \\
\Uppi_{\textbf{B}}^{(n)}(p) \enskip  & = & g^4 \; \sum_i T(R_i)\ C(R_i)\ \left( 1 - \frac{1}{2}\frac{C(G)}{C(R_i)}\right) n_i \; \frac{K^{n}}{(4 \pi^2)^2} A^{(n)}_\textbf{B}(p)\,, \\
\Uppi_{\textbf{C}}^{(n)}(p) \enskip  & = & g^2 \; C(G)\;  \frac{K^{n}}{(4 \pi^2)} A^{(n)}_\textbf{C}(p)\,.
\end{eqnarray}
It's crucial to note that the loop factors do not depend on the \emph{irrep} of the fermions. Furthermore, a sum on the fermion species can be introduced in the third contribution by inserting  
\begin{equation}
\Uppi_{\textbf{C}}^{(n)}(p) \enskip   =  g^2 \; \sum_i \frac{n_i T(R_i)}{N} C(G)\;  \frac{K^{n}}{(4 \pi^2)} A^{(n)}_\textbf{C}(p)\,.
\end{equation}
The total result can thus be written as:
\begin{multline}
\sum_n 2\Uppi_{\textbf{A}}^{\left(n\right)} + \Uppi_{\textbf{B}}^{\left(n\right)} + \Uppi_{\textbf{C}}^{\left(n\right)} = \frac{K}{N} \sum_i  \left\{ \sum_n \frac{n_i T(R_i) C(R_i)}{N} K^n \left[ 2 A^{(n-1)}_\textbf{A}(p) + A^{(n-1)}_\textbf{B}(p) \right] + \right. \\
\left. \sum_n \frac{n_i T(R_i) C(G)}{N}\ K^{n-1} \left[ A^{(n-1)}_\textbf{C}(p) - K A^{(n-1)}_\textbf{B}(p)/2 \right] \right\}\,.
\end{multline}
We can now identify again the sums leading to the abelian $F(K)$ and non-abelian $H(K)$ functions, defined in Eq.~\eqref{eq:FH}.
The $\beta$-function can thus be expressed as
\begin{equation}
    \beta (K) =  \frac{2K^2}{3} \left[ 1 + \frac{C( G)}{N}\left\{ \frac{-11}{4} + H(K)\right\} + \frac{1}{N} \left(\sum_i \frac{T( R_i)\ C(R_i)\ n_i}{N} \right) F(K)  \right]\,.
    \label{eq:beta_multiple}
\end{equation}
The coefficient in front of $F(K)$ evaluates to an $\mathcal{O} (1)$ number as it sums the degrees of freedom of the fermions weighted by the Casimirs, thus the second term is genuinely an $\mathcal{O} (1/N)$ contribution. Nevertheless, we see that for non-abelian gauge groups, the singularity driving the UV fixed point is the same as in the case of a single large-multiplicity \emph{irrep}.
Note how this result compares to Eq.~\eqref{eq:beta_semisimple}.

This result proves that  models with multiple \emph{irreps} have the same dynamics as models with a single \emph{irrep} in the large-$N_f$ limit.

\subsection{Semi-simple Gauge Group}

We now consider the most general case of a semi-simple gauge group $\mathcal{G} = \times_\alpha \mathcal{G}_\alpha$ with $n_i$ fermions $\Psi_i$ in the \emph{irrep} $R_i = \times_\alpha R_{i\alpha}$. Combining the definitions in the previous sections, we can define a fermion multiplicity for each gauge group as follows:
\begin{equation}
N_\alpha = \sum_i n_i \ \left( \Pi_{\beta \neq \alpha} d(R_{i\beta}) \right) \ T(R_{i\alpha})\,, \qquad K_\alpha = \frac{g_\alpha^2}{4 \pi^2} N_\alpha\,. 
\end{equation}
It is convenient to define effective fermion multiplicities that count the multiplicity of fermion specie $\Psi_i$ relative to one or two gauge groups respectively:
\begin{equation}
\tilde{n}_{i\alpha} = n_i \ \left( \Pi_{\beta \neq \alpha} d(R_{i\beta}) \right)\,, \quad \tilde{n}_{i\alpha\beta} = n_i \ \left( \Pi_{\gamma \neq \alpha, \beta} d(R_{i\gamma}) \right)\,,
\end{equation}
so that $N_\alpha = \sum_i \tilde{n}_{i\alpha} T(R_\alpha)$.

The generalisation of the $\beta$-function is now straightforward, starting from Eqs~\eqref{eq:beta_semisimple} and~\eqref{eq:beta_multiple}:
\begin{multline}
\beta (K_\alpha) = \frac{2 K_\alpha^2}{3} \left[ 1 + \frac{C(G_\alpha)}{N_\alpha} \left\{ -\frac{11}{4} + H(K) \right\} + \right. \\
\left. \frac{1}{N_\alpha} \left\{ \sum_i \frac{\tilde{n}_{i\alpha} T(R_{i\alpha}) C(R_{i\alpha})}{N_\alpha} F(K_\alpha) + \sum_{\beta \neq \alpha} \sum_i \frac{\tilde{n}_{i\alpha\beta} T(R_{i\alpha}) C(R_{i\beta})}{N_\beta} F(K_\beta) \right\} \right]\,.
\end{multline}
The above result confirms that the UV dynamics of the theory is fully determined by the non-abelian structure, as the additional terms proportional to $F(K)$ remain finite for $K<3$ in the case of non-abelian gauge groups.

For completeness and reference, in Appendix~\ref{app:yuk} we provide the $\beta$-functions for Yukawa couplings 
from Ref~\cite{Antipin:2018zdg}, adapted to our formalism and extended to the case of multiple \emph{irreps}.

\section{Applications} \label{sec4}

\subsection{Chiral gauge theories: generalised Georgi-Glashow and Bars-Yankielowicz models}

Chiral gauge theories have received substantial attention in the literature, especially when asymptotically free because of the interesting low energy dynamics~\cite{Raby:1979my,Georgi:1981mh}. The simplest incarnations consist of two different species of fermions, whose multiplicities are chosen to cancel the gauge anomaly. Theories of this class can be constructed on a simple SU($N_c$) gauge group, with one fermion in the symmetric or anti-symmetric \emph{irrep} and a suitable number of conjugate fundamental to cancel the gauge anomaly. They go under the names of Bars-Yankielowicz (BY)~\cite{Bars:1981se} and generalised Georgi-Glashow (GG)~\cite{Appelquist:1999vs} theories. Their low energy dynamics is still not fully understood in the asymptotic free case with small number of chiral families~\cite{Bolognesi:2019wfq,Bolognesi:2017pek}.

In this work we are interested in the limit of large number of fermions, so that we will consider a case with $n_g$ chiral generations. The fermion content of the two template theories thus consists of:
\begin{eqnarray}
\mbox{BY} & \Rightarrow & R_1 = \tiny{\yng(2)}\;\;\; [n_g]\,, \qquad R_2 = \overline{\tiny{\yng(1)}}\;\;\; [(N+4) n_g] \,; \\
\mbox{GG} & \Rightarrow & R_1 =\; \tiny{\yng(1,1)}\;\;\;\; [n_g]\,, \qquad R_2 = \overline{\tiny{\yng(1)}}\;\;\; [(N-4) n_g] \,.
\end{eqnarray}
Following Eq.~\eqref{eq:Ndef}, we can define the following fermion multiplicity:
\begin{equation}
N_{\rm BY/GG} = n_g \frac{N\pm 3}{2}\,,
\end{equation}
which is large when $n_g$ is large. Thus, the $\beta$-functions from Eq.~\eqref{eq:beta_multiple} read
\begin{equation}
\beta(K)_{\rm BY/GG} = \frac{2 K^2}{3} \left[ 1 + \frac{2 N}{n_g (N\pm 3)} \left\{ -\frac{11}{4} + H(K) \right\} + \frac{1}{n_g} \frac{3 N^2 \pm N - 4}{2 N (N\pm 3)} F(K) \right]\,,
\end{equation}
where the large-$n_g$ expansion is evident, and we recall that $N \geq 3$ for BY and $N \geq 5$ for GG.
The UV dynamics of the theories, therefore, is determined by the $H(K)$ dependence, which produces an attractive fixed point  for $K^\ast = 3$, corresponding to
\begin{equation}
\frac{g_\ast^2}{4 \pi} = \frac{6 \pi}{n_g (N\pm 3)}\,.
\end{equation}

\subsection{Split Extended GUTs}

The idea of a Grand Unification Theory (GUT) where the Standard Model is embedded in a simple gauge group has been a leading motivation for new physics.
Models based on SU(5) and SO(10) are the most minimal ones, as they can accommodate a complete standard family of fermions in a single set of chiral fermions, with or without a right-handed neutrino. Nevertheless, there has been some interest in building extended theories, where the fermion multiplets in the GUT also contain additional fermions. By construction, they are not chiral, thus they can receive a large mass from the breaking of the GUT gauge symmetry (\emph{survival hypothesis}~\cite{Georgi:1979md,Barbieri:1979ag}).

In this section, we will explore the possibility that the additional fermions do acquire a mass much smaller than the GUT symmetry breaking scale. Thus, there will be a new phase, between the SM and the GUT scale, where many fermions in addition to the standard chiral ones are present. What is the dynamics of the theory in such a phase? 
In the following, we will analyse two scenarios, in one of which the large-$N_f$ resummation can be employed.

\subsubsection{SU(9) model with 3 chiral families}

One of the main motivation for studying extended GUTs is the possibility to include the 3 standard families in a natural way, or even be able to predict their multiplicity.
In Ref.~\cite{Frampton:1979fd}, P.Frampton and S.Nandi proposed one such model based on SU(9) that, once broken down to SU(5), leads to a minimal GUT with three chiral generations.
Between the SU(5) and the SU(9) breaking scales, there may be many massive fermions, transforming as \emph{irreps} of SU(5), which may drive the SU(5) gauge coupling towards a Landau pole before the onset of the SU(9) symmetry.

The simplest realistic model, minimal in terms of the fermionic degrees of freedom and allowing all the standard Yukawa couplings, contains the following fermion \emph{irreps}:
\begin{equation}
\sum_i \Psi_{9,i} = [\textit{2}]_{36} \oplus [\textit{3}]_{84} \oplus [\textit{4}]_{126} \oplus [\textit{6}]_{84} \oplus 10 \times [\textit{8}]_{9}\,,
\end{equation}
where $[m]_d$ indicates a totally anti-symmetric $m$-index \emph{irrep} of SU(9), with dimension $d$.
We can first analyse the dynamics of the SU(9) theory, by using the formalism developed in the previous sections (even though a large-$N_f$ resummation may not be feasible). We can thus define:
\begin{equation}
N_9 = \sum_i n_i T(R_i) = \frac{47}{2}\,.
\end{equation}
This effective fermion multiplicity is large, but may not be large enough to justify the resummation~\cite{Antipin:2017ebo} once compared to the number of colours, $N_c = 9$. Yet, we can use Eq.~\eqref{eq:beta_multiple} to extract the $\beta$-function at one loop:
\begin{equation}
\beta (K_9)_{\rm SU(9)}^{\rm 1-loop} = \frac{2 K_9^2}{2} \left[ 1 + \frac{18}{47} \left( - \frac{11}{4} \right) \right] =  \frac{2 K_9^2}{2} \left( - \frac{5}{94} \right)\,,
\end{equation}
showing that the theory is asymptotic free. Next we want to study the dynamics of the SU(5) subgroup:
upon breaking SU(9) to SU(5), the matter fields decompose as
\begin{equation}
\sum_i \Psi_{5,i} = 3 \times ({\bf 10} + {\bf \bar{5}}) \oplus 14 \times ({\bf 5} + {\bf \bar{5}}) \oplus 9 \times ({\bf 10} + {\bf \bar{10}}) \oplus 55 \times {\bf 1}\,.
\end{equation}
Besides the 3 chiral families, therefore, the model contains a large number of non-chiral fermions in the fundamental and 2-index anti-symmetric \emph{irreps}. Explicit calculation shows that the fermion multiplicity for SU(5) is $N_5 = N_9 = 47/2$, which is now sufficiently larger than $N_c = 5$.
Employing the resummation in Eq.~\eqref{eq:beta_multiple}, we obtain:
\begin{equation}
\beta (K_5)_{\rm SU(5)} = \frac{2 K_5^2}{3} \left[ 1 + \frac{10}{47} \left\{ -\frac{11}{4} + H(K_5) \right\} + \frac{1506}{11045} F(K_5) \right]\,.
\end{equation}
The SU(5) theory with light non-chiral fermions flows in the UV towards an interacting fixed point and does not develop any dangerous Landau pole. In this model, therefore, one can split the fermion masses from the SU(9) breaking scale, without any limit on the mass split.

\subsubsection{$E_6$ and SU(6) models}

The only exceptional group $E_6$ that has complex \emph{irreps}, i.e. the fundamental $\bf \it 27$, is an optimal candidate for Grand Unification, as it is the only group after SO(10) that can be completely broken down to the Standard Model via fermion operators~\cite{Barbieri:1980vc}. While this possibility has been finally ruled out~\cite{Kugo:1994qr}, it remains a prime candidate of models motivated by string constructions.
One family of Standard fermions is embedded into the fundamental that, upon breaking to SU(5), decomposes as
\begin{equation}
{\bf \it 27} \to {\bf 10} \oplus {\bf \bar{5}} \oplus {\bf 5} \oplus {\bf \bar{5}} \oplus {\bf 1}\,.
\end{equation}
Thus, the minimal $E_6$ model contains only 3 Dirac $\bf 5$'s: the effective fermion multiplicity reads $N_5 = 9/2$, and the theory remains asymptotic free.

SU(6) has also been proposed as a candidate~\cite{Fukugita:1981gn,Hartanto:2005jr} because of interesting model building in the neutrino sector. 
One Standard family is thus embedded in the following representations:
\begin{equation}
[\textit{2}]_{15} \oplus 2 \times [\textit{5}]_{6}\,,
\end{equation}
which gives, upon breaking SU(6) to SU(5) to
\begin{equation}
{\bf 10} \oplus 2 \times {\bf \bar{5}} \oplus {\bf 5} \oplus {\bf 1}\,,
\end{equation}
thus leading to the same field content of the $E_6$ model.

\section{Conclusions} \label{sec5}

Large-$N_f$ resummation has proven a useful tool to study the UV dynamics of gauge theories with large multiplicity of fermions. This has lead to the conjecture of the emergence of an attractive interacting fixed point in the UV, which is due to the presence of a pole in the resummed leading order in the beta function.
In this note, we extended the standard formalism to include cases with multiple fermion representations. We showed that the pole can be traced back to the intrinsic non-abelian nature of the gauge group, independently on the specific representations of the fermions. Thus, we support the conjecture for non-abelian gauge groups.

As a consequence, the pole, and the UV fixed point, are also found in theories with multiple representations. We apply these results to the simplest cases of chiral gauge theories, based on SU($N_c$) gauge groups. As long as a large-enough number of chiral families are added, the theories develop an UV safe dynamics. As an example with physical interest, we study extended Grand Unified Theories, where a large number of fermions may be allowed below the symmetry breaking scale. We show, in one specific example based on SU(9) without supersymmetry, that the mass of such fermions can be split from the symmetry breaking scale arbitrarily, because the low energy SU(5) GUT flows towards a UV fixed point. Thus, the onset of SU(9) can be delayed to arbitrarily large energies.

\end{fmffile}

\section*{Acknowledgements}

The authors acknowledge partial support from the France-China Particle Physics Lab (FCPPL) and the Labex-LIO (Lyon Institute of Origins) under grant ANR-10-LABX-66 (Agence Nationale pour la Recherche), and FRAMA (FR3127, F\'ed\'eration de Recherche ``Andr\'e Marie Amp\`ere'').

\appendix

\section{Large-$N_f$ $\beta$-functions for Yukawa and quartic couplings} \label{app:yuk}

For completeness, we will list here the resummed $\beta$-functions for Yukawa couplings 
from Ref.~\cite{Antipin:2018zdg}, adapted to the notation we use in this paper, and to the case with multiple fermion \emph{irreps}.
We complete the model by adding a scalar field $\phi^A$ in the \emph{ irrep} $R_\phi = \times_\alpha R_{\phi \alpha}$, where $A$ is a generic gauge index for the scalar \emph{irrep}. The new Lagrangian interactions read:
\begin{equation}
    \mathcal{L} \supset - y_{A\mathsf{b} \mathfrak{c}} \; \phi^A \; \Psi_i^{\mathsf{b}} \; \Psi_j^{\mathfrak{c}} - \lambda^{AB}_{CD}\; \phi^{\ast, A} \phi^{\ast, B} \phi_C \phi_D + \mbox{h.c.} \,,
\end{equation}
where $\mathsf{b}$ and $\mathfrak{c}$ are gauge indices of the fermion \emph{irreps}, and it is understood that the scalar gauge quantum numbers allow for the Yukawa coupling to be gauge-invariant. First we note that the scalar will contribute to the gauge $\beta$-function with a $1/N_f$ term
\begin{equation}
   \left. \Delta \beta(K_\alpha) \right|_{\rm scalar} \supset \frac{2K^2_\alpha}{3} \frac{T\left(R_{\phi \; \alpha} \right)}{ 4 N_\alpha}\,,
\end{equation}
which corresponds to the one-loop result.

We recall that consistency of the expansion requires that the Yukawa and quartic couplings shall scale in a give way with $N_f$, as follows:
\begin{equation}
y \sim \frac{1}{\sqrt{N_f}}\,. 
\end{equation} 
This counting ensures that the non-gauge contribution to the beta function at one loop counts $1/N_f$ in the expansion.
The $\beta$-function for the Yukawa coupling reads:
\begin{align*}
    \beta\left( y_{A\mathsf{b}\mathfrak{c}} \right) =& \frac{1}{32 \pi^2} \left[ \left( y_D y^{\ast,D} y_A  \right)_{\mathsf{b}\mathfrak{c}} + \left( y_A y^{\ast, D} y_D  \right)_{\mathsf{b}\mathfrak{c}} + 2 \mathrm{Tr}\left[ y_A y^{\ast, D} \right] y_{D\mathsf{b}\mathfrak{c}} \right] \\
    & - y_{A\mathsf{b}\mathfrak{c}}\sum_{\alpha} \frac{3 K_\alpha}{4 N_\alpha} H_0\left( \frac{2K_\alpha}{3} \right) \left[ C\left( R_{i \; \alpha}\right) + C\left( R_{j \; \alpha} \right) + \frac{K_\alpha}{6} C\left( R_{\phi \; \alpha }\right)\right]\,,
\end{align*}
where the traces and contractions are intended for the fermion gauge indices, and
\begin{equation}
H_0(x)= \frac{\left( 1- \frac{x}{3}\right)\Gamma\left( 4-x \right)}{3 \Gamma^2\left( 2 -\frac{x}{2} \right) \Gamma\left( 3 -\frac{x}{2}  \right)\Gamma\left( 1 +\frac{x}{2}\right)}\,.
\end{equation}
We recall that $H_0$ is a smooth function up to $K = 15/2$, where a pole is developed. Thus, for non-abelian semi-simple groups, the gauge contribution to the Yukawa running remains finite. If it dominates over the contribution of the pure Yukawa term, the coupling will flow towards a non-interactive fixed point (asymptotic free).

%
%

\bibliographystyle{JHEP-2-2}
\bibliography{biblio}

\end{document}